# The Zeta Function for the Triangular Potential

M.G. Naber
mnaber@monroeccc.edu
Department of Science and Mathematics
Monroe County Community College
1555 S. Raisinville Rd
Monroe, Michigan, 48161-9746
11 September 2021.  Revised 23 November 2021

The zeta functions for the Schrödinger equation with a triangular potential are investigated.  Values of the zeta functions are computed using both the Weierstrass factorization theorem and analytic continuation via contour integration.  The results were found to be consistent where the domains of the two methods overlap.  Analytic continuation is used to compute values of the zeta functions at zero and the negative integers, explore the pole structure (and residue values), as well as the value of the slopes at the origin.  Those results are used for the computation of the trace and determinant of the associated Hamiltonians.

## I. INTRODUCTION

In this paper the zeta functions for the energy eigenvalues of the Schrödinger equation with a potential that is proportional to the absolute value of the spatial coordinate are considered. The goal of this paper is to explore the properties of the associated zeta functions and to compute the determinant of the Hamiltonians.  In the literature this potential is commonly referred to as a triangular potential well[1], motion in a homogeneous field[2], or sometimes a quantum bouncer[3] (which produces only odd states).  The Schrödinger equation for this potential takes the form of,

$$\left(-\frac{\hbar^2}{2\mu}\partial_z^2 + a|z|\right)\psi = E\psi, \qquad (1)$$

with appropriate boundary and domain conditions.

Zeta functions are a convenient means of computing traces and determinants of differential operators[4-7].  The trace can be used to compute Casimir effects, vacuum energies (cf. chapter 4 of Ref. 4 and chapter 5 of Ref. 5) and the determinant can be used to compute one-loop graphs in quantum field theory (page 8 of Ref. 5).  Given a set of positive numbers, $\lambda_n$, where $\lambda_1 > 0$ and $\lambda_{n+1} > \lambda_n$ for all $n$, a zeta function can be defined as[7],

$$\zeta(s) = \sum_{n=1}^{\infty} \lambda_n^{-s}, \qquad (2)$$

where $s$ is taken to be a complex number whose real part is sufficiently large so that the series is convergent.  Analytic continuation in the variable $s$ can be used to assign a finite value to the above sum where it would otherwise be considered divergent (cf. chapter 1 of Ref. 5).  The above definition is of course a generalization of the well-known Riemann zeta function (cf. chapter 1 of Ref. 4),

$$\zeta_R(s) = \sum_{n=1}^{\infty} n^{-s}. \tag{3}$$

If the spectrum of a differential operator is used to define the zeta function, the trace and determinant of the differential operator can be computed using the following formulae (cf. sec. 1.5 of Ref. 5 and sec. 2.1 of Ref. 6). Let $H$ be a differential operator with spectrum $\lambda_n$, denote by $g_n$ the multiplicity of $\lambda_n$, and eigenfunctions $y_n$ (i.e., $Hy_n = \lambda_n y_n$ with suitable boundary conditions) then the zeta function associated with $H$ is given by,

$$\zeta_H(s) = \sum_{n=1}^{\infty} g_n \lambda_n^{-s}. \tag{4}$$

For the potential studied in this paper none of the eigenvalues are degenerate ($g_n = 1$). The trace of the differential operator is then,

$$tr(H) = \sum_{n=1}^{\infty} \lambda_n = \zeta_H(-1), \tag{5}$$

and the determinant is given by,

$$det(H) = \prod_{n=1}^{\infty} \lambda_n = e^{-\frac{d}{ds}\zeta_H(s)|_{s=0}}. \tag{6}$$

In section II the solutions for the Schrödinger equation are given, a discussion of the partition of the problem into odd and even states, and how the energy eigenvalues for each case can be obtained. In section III the Weierstrass factorization theorem[8] is used to work out some special values for the zeta functions. Typographical errors that exist in the literature are also pointed out. In section IV contour integration will be used to find a representation for the zeta functions that will show where the poles are, the values of the associated residue, and provide an alternate means of computing special values of the zeta functions. The section ends with the computation of the determinant of the Hamiltonians and the results are compared with existing results in the literature.

## II. HAMILTONIANS FOR THE PROBLEM

The Schrödinger equation for the triangular potential is even in the independent variable $z$,

$$\left(\partial_z^2 + \frac{2\mu}{\hbar^2}(E - a|z|)\right)\psi = 0. \tag{7}$$

Hence, the solutions will be odd or even with respect to the independent variable. The odd solutions will obey $\psi_{odd}(0) = 0$, while the even solutions will obey, $\partial_z \psi_{even}(0) = 0$[1]. As the even and odd states have different "boundary conditions" two different zeta functions will be constructed. For the odd states,

$$\left(-\frac{\hbar^2}{2\mu}\partial_z^2 + a|z|\right)\psi = H_o\psi = E\psi, \tag{8}$$

$$\psi_{odd}(0) = 0, \tag{9}$$

$$\zeta_{H_o}(s) = \sum E^{-s}, \tag{10}$$

and for the even states,

$$\left(-\frac{\hbar^2}{2\mu}\partial_z^2 + a|z|\right)\psi = H_\varepsilon\psi = E\psi, \tag{11}$$

$$\partial_z\psi_{even}(0) = 0, \tag{12}$$

$$\zeta_{H_\varepsilon}(s) = \sum E^{-s}. \tag{13}$$

Ref. 1 goes into a fair amount of detail in finding the solutions. There is also a brief discussion in Ref. 9. To find the solutions and the eigenvalues, first make the Schrödinger equation dimensionless. Let,

$$E = \mathcal{E}\left(\frac{a^2\hbar^2}{2\mu}\right)^{\frac{1}{3}}, \text{ and, } z = x\left(\frac{\hbar^2}{2\mu a}\right)^{\frac{1}{3}}. \tag{14}$$

The Schrödinger equation then becomes,

$$\partial_x^2\psi + (\mathcal{E} - |x|)\psi = 0. \tag{15}$$

Now shift the independent variable to find the Airy equation. For $x > 0$ or $y > -\mathcal{E}$, let,

$$y = x - \mathcal{E}, \tag{16}$$

$$\Rightarrow \partial_y^2\psi - y\psi = 0. \tag{17}$$

There are two linearly independent solutions to this equation, the Airy function of the first kind, $Ai(y)$, and the Airy function of the second kind, $Bi(y)$[9]. The wave function needs to go to zero at infinity, so, the desired solution is given by $Ai(y)$ as $Bi(y)$ becomes unbounded, hence,

$$\psi = C_0 Ai(y). \tag{18}$$

Denote the roots of the Airy function by $o_n$ where $n = 1,2,3, ...$ then the energy eigenvalues for the odd states are given by (recall that the roots of the Airy function are negative)[1,9]

$$\mathcal{E} = -o_n, \qquad (19)$$

$$\Rightarrow E = -o_n \left(\frac{a^2 \hbar^2}{2\mu}\right)^{\frac{1}{3}} = |o_n| \left(\frac{a^2 \hbar^2}{2\mu}\right)^{\frac{1}{3}}. \qquad (20)$$

The corresponding zeta function, for just the odd states, is then,

$$\zeta_{H_o}(s) = \left(\frac{a^2 \hbar^2}{2\mu}\right)^{-\frac{s}{3}} \sum_{n=1}^{\infty} |o_n|^{-s}. \qquad (21)$$

The roots of the Airy function are not obtainable in closed form. Asymptotically, the roots of the Airy function (or, here, the energy eigenvalues) grow as[10], $o_n \sim n^{\frac{2}{3}}$. Hence $\zeta_{H_o}(s)$ is only convergent for $\Re(s) > 3/2$, where $\Re(s)$ denotes the real part of $s$. In fact, $\zeta_{H_o}$ is analytic[10] for $\Re(s) > 3/2$. This will allow for the zeta function to be meromorphically continued throughout the complex plane[10].

The Airy function is also the solution for the even states; however, the energy eigenvalues are given by the roots of the slope of the Airy function[1,9]. Denote the roots of the first derivative of the Airy function by $\varepsilon_n$ where $n = 1,2,3, ...$ then the energy eigenvalues for the even states are given by (recall that the roots are negative),

$$\mathcal{E} = -\varepsilon_n, \qquad (22)$$

$$\Rightarrow E = -\varepsilon_n \left(\frac{a^2 \hbar^2}{2\mu}\right)^{\frac{1}{3}} = |\varepsilon_n| \left(\frac{a^2 \hbar^2}{2\mu}\right)^{\frac{1}{3}}. \qquad (23)$$

Just as in the odd case, the roots of the slope of the Airy function are not obtainable in closed form. Note that the roots of the first derivative of the Airy function are interlaced between the roots of the Airy function itself (Laguerre's theorem on separation zeros, see page 92 of Ref. 8). This means that the roots of the first derivative must also grow asymptotically like $n^{\frac{2}{3}}$. The corresponding zeta function, for just the even states, is then,

$$\zeta_{H_\varepsilon}(s) = \left(\frac{a^2 \hbar^2}{2\mu}\right)^{-\frac{s}{3}} \sum_{n=1}^{\infty} |\varepsilon_n|^{-s}. \qquad (24)$$

which is convergent for $\Re(s) > 3/2$.

## III. ZETA FUNCTION VIA WEIERSTRASS

Following Ref. 9, and keeping some of their notation, some values of the zeta functions can be computed without too much effort. Since the Airy function is an entire function, we can use a result due to Weierstrass (page 88 of Ref. 8), namely, the Weierstrass factorization theorem,

$$Ai(y) = Ai(0)e^{-\kappa y} \prod_{n=1}^{\infty}\left(1 + \frac{y}{|o_n|}\right) e^{-\frac{y}{|o_n|}}, \quad (25)$$

where,

$$\kappa = -\frac{Ai'(0)}{Ai(0)} = \frac{3^{\frac{5}{6}}\Gamma\left(\frac{2}{3}\right)^2}{2\pi}. \quad (26)$$

Eq. (26) is a known value (see page 17 of Ref. 9 or one can use MAPLE). The logarithmic derivative of Eq. (25) is

$$\frac{Ai'(y)}{Ai(y)} = -\kappa + \sum_{n=1}^{\infty} \frac{1}{|o_n| + y} - \frac{1}{|o_n|}. \quad (27)$$

Now apply $\partial_y$ and use the Airy differential equation, Eq. (17), to rearrange the result to obtain,

$$y - \left(\frac{Ai'(y)}{Ai(y)}\right)^2 = -\sum_{n=1}^{\infty} \frac{1}{(|o_n| + y)^2}. \quad (28)$$

Setting $y = 0$ gives,

$$\kappa^2 = \sum_{n=1}^{\infty} \frac{1}{|o_n|^2}, \quad (29)$$

$$\Rightarrow \zeta_{H_o}(2) = \left(\frac{2\mu}{a^2\hbar^2}\right)^{\frac{2}{3}} \kappa^2. \quad (30)$$

$\zeta_{H_o}(3)$ can be obtained by applying $\partial_y$ to Eq. (28) and suitably rearranging,

$$1 - 2\left(\frac{Ai'(y)}{Ai(y)}\right)\left(y - \left(\frac{Ai'(y)}{Ai(y)}\right)^2\right) = \sum_{n=1}^{\infty} \frac{2}{(|o_n| + y)^3}. \quad (31)$$

Now let $y = 0$,

$$\frac{1}{2} - \kappa^3 = \sum_{n=1}^{\infty} \frac{1}{|o_n|^3}, \quad (32)$$

$$\Rightarrow \zeta_{H_o}(3) = \frac{2\mu}{a^2\hbar^2}\left(\frac{1}{2} - \kappa^3\right). \tag{33}$$

Similarly, $\zeta_{H_o}(4)$ can be computed,

$$\kappa^4 - \frac{\kappa}{3} = \sum_{n=1}^{\infty} \frac{1}{|o_n|^4}. \tag{34}$$

This last result differs in sign from Ref. 9. The above result is positive which is what would be expected when summing a convergent series of positive numbers,

$$\Rightarrow \zeta_{H_o}(4) = \left(\frac{2\mu}{a^2\hbar^2}\right)^{\frac{4}{3}}\left(\kappa^4 - \frac{\kappa}{3}\right). \tag{35}$$

Clearly this procedure could be carried out indefinitely to obtain $\zeta_{H_o}(n)$ for any $n > 1$.

Using the Weierstrass construction again, but this time for the first derivative of the Airy function, will give some special values for the zeta function of the even states,

$$Ai'(y) = Ai'(0) \prod_{n=1}^{\infty} \left(1 + \frac{y}{|\varepsilon_n|}\right) e^{-\frac{y}{|\varepsilon_n|}}. \tag{36}$$

Notice that there is no leading exponential factor because the Airy function has an inflection point at the origin. In a similar fashion to the calculation for the odd states, take the logarithm of Eq. (36), compute the derivative, and rearrange using the Airy differential equation, Eq. (17),

$$\frac{yAi(y)}{Ai'(y)} = \sum_{n=1}^{\infty} \frac{1}{|\varepsilon_n| + y} - \frac{1}{|\varepsilon_n|}. \tag{37}$$

Taking the derivative once more and letting $y = 0$ gives the first result,

$$\frac{Ai(0)}{Ai'(0)} = -\sum_{n=1}^{\infty} \frac{1}{|\varepsilon_n|^2}, \tag{38}$$

or,

$$\frac{1}{\kappa} = \sum_{n=1}^{\infty} \frac{1}{|\varepsilon_n|^2}, \tag{39}$$

$$\Rightarrow \zeta_{H_\varepsilon}(2) = \left(\frac{2\mu}{a^2\hbar^2}\right)^{\frac{2}{3}} \frac{1}{\kappa}. \tag{40}$$

Taking two derivatives of Eq. (37), rearranging, and setting $y = 0$, yields,

$$1 = \sum_{n=1}^{\infty} \frac{1}{|\varepsilon_n|^3}, \tag{41}$$

$$\Rightarrow \zeta_{Ae}(3) = \frac{2\mu}{a^2\hbar^2}. \tag{42}$$

Continuing the process once more gives,

$$\frac{1}{2\kappa^2} = \sum_{p=1}^{\infty} \frac{1}{|\varepsilon_n|^4}. \tag{43}$$

Note that this result differs by a factor of 1/2 from Ref. 9. This result, and Eq. (34), will be confirmed in the next section using analytic continuation. The zeta function at $s = 4$ is then,

$$\zeta_{H_\varepsilon}(4) = \left(\frac{2\mu}{a^2\hbar^2}\right)^{\frac{4}{3}} \frac{1}{2\kappa^2}. \tag{44}$$

Clearly $\zeta_{H_\varepsilon}(n)$ can be computed in this way for any $n > 1$.

## IV. ZETA FUNCTION VIA CONTOUR INTEGRATION

There is an alternate means of computing the zeta function which will allow for its evaluation at points where the sum is not convergent, that is, by using a contour integral to analytically continue the zeta function[11,12]. Consider the odd states first (physical factor to be added later). Recall the following result from complex analysis. Let $f(z)$ have zeros at $z_1, z_2, z_3, \ldots$. Let $\gamma$ be a simple closed curve that captures all the zeros of $f(z)$, then,

$$\frac{1}{2\pi i} \oint_\gamma z^s \frac{f'(z)}{f(z)} dz = \sum_{n=1}^{\infty} (z_n)^s. \tag{45}$$

The roots of the Airy function all occur on the negative real axis. The path, $\gamma$, should not cross the negative real axis as this is a branch cut for the function $z^s$. To use the above result let,

$$f(z) = Ai(-z). \tag{46}$$

and let $\gamma$ be a closed curve that starts just to the right of the origin and just below the positive real axis, then goes out to positive infinity, crosses the real axis, and then goes just above the real axis traveling back to just before the origin, and then crosses the axis again coming back to the starting point. This contour captures all the roots of the Airy function (with the sign of the argument flipped) traveling in a counterclockwise fashion but avoids the origin and the negative real axis,

$$\frac{1}{2\pi i}\oint_\gamma z^{-s}\frac{\frac{d}{dz}Ai(-z)}{Ai(-z)}dz = \sum_{n=1}^{\infty}|o_n|^{-s}. \qquad (47)$$

The sum on the right goes over all the roots of the Airy function. If the contour is deformed in the same way as in Refs. 7, 11, and 12, i.e., the piece just above the positive real axis is lifted and folded back to be crossing the positive imaginary axis and being just above the negative real axis and the piece just below the positive real axis being pulled down and folded back to be crossing the negative imaginary axis and being just below the negative real axis. Essentially, the deformed path is a circle (whose radius will go to infinity) that starts just below the negative real axis travels on a circular arc in a counterclockwise fashion until just above the negative real axis. The path then cuts in toward the origin traveling just above the negative real axis towards the origin going around the origin (in a clockwise sense) and then traveling just below the negative real axis going back out to negative infinity. Once the radius of the circle goes to infinity the contour integral (left hand side of Eq. (47)) becomes,

$$\frac{\sin(\pi s)}{\pi}\int_0^\infty x^{-s}\frac{Ai'(x)}{Ai(x)}dx = \sum_{n=1}^{\infty}|o_n|^{-s}. \qquad (48)$$

Similarly for the first derivative of the Airy function,

$$\frac{\sin(\pi s)}{\pi}\int_0^\infty x^{-s}\frac{Ai''(x)}{Ai'(x)}dx = \sum_{n=1}^{\infty}|\varepsilon_n|^{-s}. \qquad (49)$$

To explore the pole structure and to compute the trace and the determinant of the Hamiltonian note the following two series representations for the logarithmic derivative of the Airy function, about $x = 0$ and $x = \infty$ respectively,

$$\frac{Ai'(x)}{Ai(x)} = \sum_{n=0}^{\infty} a_n x^n \qquad (50)$$

where,

$$a_0 = -\frac{3^{\frac{5}{6}}\Gamma\left(\frac{2}{3}\right)^2}{2\pi} = -\kappa, \ a_1 = -\frac{3^{\frac{5}{3}}\Gamma\left(\frac{2}{3}\right)^4}{4\pi^2} = -\kappa^2, \ a_2 = \frac{1}{2} - \frac{3^{\frac{5}{2}}\Gamma\left(\frac{2}{3}\right)^6}{8\pi^3} = \frac{1}{2} - \kappa^3, \ \ldots \ (51)$$

and,

$$\frac{Ai'(x)}{Ai(x)} = \sqrt{x}\left(\sum_{n=1}^{\infty}(-1)^n b_n\left(x^{-\frac{3}{2}}\right)^n - 1\right), \qquad (52)$$

where,

$$b_1 = \frac{1}{4}, \ b_2 = \frac{5}{32}, \ b_3 = \frac{15}{64}, \ \ldots \qquad (53)$$

Higher order coefficients are readily available via MAPLE or similar symbolic computation software. For convenience, adopt the following notation for the zeta function for the odd states,

$$\zeta_{H_o}(s) = \alpha^{-s} f(s), \tag{54}$$

where,

$$\alpha = \left(\frac{a^2 \hbar^2}{2\mu}\right)^{\frac{1}{3}}, \tag{55}$$

and,

$$f(s) = \frac{\sin(\pi s)}{\pi} \int_0^\infty x^{-s} \frac{Ai'(x)}{Ai(x)} dx, \tag{56}$$

The trace and determinant in this notation are then,

$$tr(H) = \alpha f(-1), \tag{57}$$

$$det(H) = \alpha^{f(0)} e^{-\frac{d}{ds} f(s)|_{s=0}}. \tag{58}$$

To evaluate $f(s)$, Eq. (56), the integral needs to be split at $x = 1$ and then substitute in the two series representations for the logarithmic derivative of the Airy function. This is what is done when evaluating the integral representation of the Riemann and other zeta functions (see Ref. 4 for similar examples),

$$f(s) = \frac{\sin(\pi s)}{\pi} \left( \int_0^1 x^{-s} \frac{Ai'(x)}{Ai(x)} dx + \int_1^\infty x^{-s} \frac{Ai'(x)}{Ai(x)} dx \right), \tag{59}$$

$$f(s) = \frac{\sin(\pi s)}{\pi} \left( \sum_{n=0}^\infty a_n \int_0^1 x^{n-s} dx - \int_1^\infty x^{-s+\frac{1}{2}} dx + \sum_{n=1}^\infty (-1)^n b_n \int_1^\infty x^{-s+\frac{1}{2}-\frac{3}{2}n} dx \right). \tag{60}$$

To analytically continue $f(s)$ each of the three integrals is treated separately: to evaluate the first integral take $\Re(s) < 1$, to evaluate the second integral take $\Re(s) > 3/2$, and finally, for the third integral, take $\Re(s) > 0$. This will avoid logarithms. Combining all three results gives,

$$f(s) = \frac{\sin(\pi s)}{\pi} \left( \sum_{n=0}^\infty \frac{a_n}{n+1-s} + \frac{1}{\frac{3}{2}-s} - \sum_{n=1}^\infty \frac{(-1)^n b_n}{\frac{3}{2}(1-n)-s} \right). \tag{61}$$

Expressed in this form the poles are easily identifiable and, they are all simple. Note that,

$$\frac{\sin(\pi s)}{\pi} \to s \quad \text{as } s \to 0, \tag{62}$$

$$\frac{\sin(\pi s)}{\pi} \to -(s-1) \quad \text{as } s \to 1, \tag{63}$$

$$\frac{\sin(\pi s)}{\pi} \to (s-2) \quad \text{as } s \to 2, \tag{64}$$

$$etc.,$$

i.e., $\sin(\pi s)$ will go to zero as $s$ approaches any integer. So, we will not get a pole when $s$ is an integer. The first sum in Eq. (61) will not generate any poles. We will get a pole when

$$s = \frac{3}{2}, -\frac{3}{2}, -\frac{9}{2}, \ldots, -\frac{3}{2}(2k-1) \quad \text{for } k = 0, 1, 2, \ldots. \tag{65}$$

These are the same locations as were found in Ref. 10, which used an alternate method. The residue at $s = -3(2k-1)/2$ is then seen to be proportional to $b_{2k}$ (take $b_0 = 1$).

Eq. (61) can be used to compute $f(0)$ by the following limit,

$$f(0) = \lim_{s \to 0} \frac{\sin(\pi s)}{\pi} \left( \sum_{n=0}^{\infty} \frac{a_n}{n+1-s} + \frac{1}{\frac{3}{2}-s} - \sum_{n=1}^{\infty} \frac{(-1)^n b_n}{\frac{3}{2}(1-n)-s} \right). \tag{66}$$

As $a_n$ forms a convergent series, $a_n/(n+1)$ also forms a convergent series. The only term in the second series that survives the limit is the first term,

$$f(0) = \lim_{s \to 0} s \left[ \sum_{n=0}^{\infty} \frac{a_n}{n+1} + \frac{2}{3} - \frac{b_1}{s} \right], \tag{67}$$

$$\Rightarrow f(0) = -b_1 = -\frac{1}{4}. \tag{68}$$

$f(s)$ can be evaluated for the negative integers as well. Consider the following limit,

$$f(-1) = \lim_{s \to -1} \frac{\sin(\pi s)}{\pi} \left( \sum_{n=0}^{\infty} \frac{a_n}{n+1-s} + \frac{1}{\frac{3}{2}-s} - \sum_{n=1}^{\infty} \frac{(-1)^n b_n}{\frac{3}{2}(1-n)-s} \right), \tag{69}$$

$$f(-1) = \lim_{t \to 0} -t \left( \sum_{n=0}^{\infty} \frac{a_n}{n+2-t} + \frac{1}{\frac{5}{2}-t} - \sum_{n=1}^{\infty} \frac{(-1)^n b_n}{\frac{5}{2}-\frac{3}{2}n-t} \right), \tag{70}$$

$$f(-1) = 0, \quad \Rightarrow tr(H_o) = 0. \tag{71}$$

Similarly, for $s = -2$, $f(-2) = 0$. For $s = -3$ the limit needs to be examined,

$$f(-3) = \lim_{s \to -3} \frac{sin(\pi s)}{\pi} \left( \sum_{n=0}^{\infty} \frac{a_n}{n+1-s} + \frac{1}{\frac{3}{2}-s} - \sum_{n=1}^{\infty} \frac{(-1)^n b_n}{\frac{3}{2}(1-n)-s} \right), \tag{72}$$

$$f(-3) = \lim_{t \to 0} -t \left[ \sum_{n=0}^{\infty} \frac{a_n}{n+4-t} + \frac{1}{\frac{9}{2}-t} - \sum_{n=1}^{\infty} \frac{(-1)^n b_n}{\frac{3}{2}(1-n)+3-t} \right]. \tag{73}$$

Only the $n = 3$ term in the second series will survive the limit,

$$f(-3) = \lim_{t \to 0} t \frac{-b_3}{t} = -b_3 = -\frac{15}{64}. \tag{74}$$

The above pattern repeats, $f(-4) = f(-5) = 0$, and $f(-6) = b_5$. In general, for the non-zero results,

$$f(-3n) = (-1)^n b_{2n+1} \quad \text{for } n = 1,2,3,\dots. \tag{75}$$

It is interesting to note that only $b_n$ with odd indices appear in the above formula while the even indexed $b_n$ appear in the residue.

Similarly, $f(s)$ can also be evaluated for the positive integers. Consider the following,

$$f(1) = \lim_{s \to 1} \frac{sin(\pi s)}{\pi} \left( \sum_{n=0}^{\infty} \frac{a_n}{n+1-s} + \frac{1}{\frac{3}{2}-s} - \sum_{n=1}^{\infty} \frac{(-1)^n b_n}{\frac{3}{2}(1-n)-s} \right), \tag{76}$$

$$f(1) = \lim_{t \to 0} -t \left( \sum_{n=0}^{\infty} \frac{a_n}{n-t} + \frac{1}{\frac{1}{2}-t} - \sum_{n=1}^{\infty} \frac{(-1)^n b_n}{\frac{3}{2}(1-n)-1-t} \right). \tag{77}$$

Only the $n = 0$ term in the first series will survive the limit,

$$\Rightarrow f(1) = a_0. \tag{78}$$

For $s = 2$,

$$f(2) = \lim_{s \to 2} \frac{sin(\pi s)}{\pi} \left( \sum_{n=0}^{\infty} \frac{a_n}{n+1-s} + \frac{1}{\frac{3}{2}-s} - \sum_{n=1}^{\infty} \frac{(-1)^n b_n}{\frac{3}{2}(1-n)-s} \right), \tag{79}$$

$$f(2) = \lim_{t \to 0} t \left( \sum_{n=0}^{\infty} \frac{a_n}{n-1-t} + \frac{1}{\frac{3}{2}-2-t} - \sum_{n=1}^{\infty} \frac{(-1)^n b_n}{\frac{3}{2}(1-n)-2-t} \right). \tag{80}$$

Only the $n = 1$ term in the first series will survive the limit,

$$\Rightarrow f(2) = -a_1. \tag{81}$$

In general,

$$f(n+1) = (-1)^n a_n \quad \text{for } n = 0,1,2,\ldots. \tag{82}$$

Note that this gives the same results that were found in section III using the Weierstrass factorization method.

The previous computations can be repeated for the even order solutions. Define $g(s)$ analogously to $f(s)$, Eq. (56),

$$g(s) = \sum_{n=1}^{\infty} |\varepsilon_n|^{-s} = \frac{\sin(\pi s)}{\pi} \int_0^{\infty} x^{-s} \frac{Ai''(x)}{Ai'(x)} dx. \tag{83}$$

Again, expand $Ai''(x)/Ai'(x)$ about the origin and infinity,

$$\frac{Ai''(x)}{Ai'(x)} = \sum_{n=1}^{\infty} c_n x^n, \tag{84}$$

where,

$$c_1 = -\frac{2\pi}{3^{\frac{5}{6}}\Gamma\left(\frac{2}{3}\right)^2}, \quad c_2 = 1, \quad c_3 = -\frac{2\pi^2}{3^{\frac{5}{3}}\Gamma\left(\frac{2}{3}\right)^4}, \quad c_4 = \frac{4\pi}{3^{\frac{11}{6}}\Gamma\left(\frac{2}{3}\right)^2}, \quad \ldots, \tag{85}$$

and,

$$\frac{Ai''(x)}{Ai'(x)} = -\sqrt{x} - \sqrt{x} \sum_{n=1}^{\infty} (-1)^n d_n \left(x^{\frac{3}{2}}\right)^{-n}, \tag{86}$$

where,

$$d_1 = \frac{1}{4}, \quad d_2 = \frac{7}{32}, \quad d_3 = \frac{21}{64}, \quad \ldots. \tag{87}$$

Split the integral, Eq. (83), at $x = 1$ and follow the same steps as were used for the evaluation of $f(s)$,

$$g(s) = \frac{sin(\pi s)}{\pi} \left( \int_0^1 x^{-s} \frac{Ai''(x)}{Ai'(x)} dx + \int_1^\infty x^{-s} \frac{Ai''(x)}{Ai'(x)} dx \right), \tag{88}$$

$$g(s) = \frac{sin(\pi s)}{\pi} \left( \sum_{n=1}^\infty c_n \int_0^1 x^{n-s} dx - \int_1^\infty x^{-s+\frac{1}{2}} dx - \sum_{n=1}^\infty (-1)^n d_n \int_1^\infty x^{-s+\frac{1}{2}-\frac{3}{2}n} dx \right). \tag{89}$$

To complete the analytic continuation of $g(s)$ each of the three integrals is treated separately: to evaluate the first integral take $\Re(s) < 1$, to evaluate the second integral take $\Re(s) > 3/2$, and finally, for the third integral, take $\Re(s) > 0$. Combining all three results gives,

$$g(s) = \frac{sin(\pi s)}{\pi} \left( \sum_{n=1}^\infty \frac{c_n}{n+1-s} + \frac{1}{\frac{3}{2}-s} + \sum_{n=1}^\infty \frac{(-1)^n d_n}{\frac{3}{2}(1-n)-s} \right). \tag{90}$$

Just as for $f(s)$, the poles are easily identifiable, and are all simple, occurring at

$$s = \frac{3}{2}, -\frac{3}{2}, -\frac{9}{2}, \ldots, -\frac{3}{2}(2k-1) \quad \text{for } k = 0,1,2,\ldots. \tag{91}$$

Note that the residue at $s = -3(2k-1)/2$ is proportional to $d_{2k}$, (take $d_0 = 1$).

$g(s)$ can also be readily evaluated at integer values of $s$. First consider the origin,

$$g(0) = \lim_{s \to 0} \frac{sin(\pi s)}{\pi} \left( \sum_{n=1}^\infty \frac{c_n}{n+1-s} + \frac{1}{\frac{3}{2}-s} + \sum_{n=1}^\infty \frac{(-1)^n d_n}{\frac{3}{2}(1-n)-s} \right), \tag{92}$$

$$g(0) = \lim_{s \to 0} s \left( \sum_{n=1}^\infty \frac{(-1)^n d_n}{\frac{3}{2}(1-n)-s} \right). \tag{93}$$

Only the first term in the series will survive the limit,

$$\Rightarrow g(0) = \lim_{s \to 0} s \left( -d_1 \frac{1}{-s} \right) = d_1 = \frac{1}{4}. \tag{94}$$

The positive integers generate a similar pattern for $g(s)$ as they did for $f(s)$. For $s = 1$,

$$g(1) = \lim_{s \to 1} \frac{sin(\pi s)}{\pi} \left( \sum_{n=1}^\infty \frac{c_n}{n+1-s} + \frac{1}{\frac{3}{2}-s} + \sum_{n=1}^\infty \frac{(-1)^n d_n}{\frac{3}{2}(1-n)-s} \right), \tag{95}$$

$$g(1) = \lim_{t \to 0} -t \left( \sum_{n=1}^{\infty} \frac{c_n}{n-t} + \frac{1}{\frac{1}{2}-t} + \sum_{n=1}^{\infty} \frac{(-1)^n d_n}{\frac{3}{2}(1-n)-1-t} \right), \quad (96)$$

$$\Rightarrow g(1) = 0. \quad (97)$$

For $s = 2$,

$$g(2) = \lim_{s \to 2} \frac{\sin(\pi s)}{\pi} \left( \sum_{n=1}^{\infty} \frac{c_n}{n+1-s} + \frac{1}{\frac{3}{2}-s} + \sum_{n=1}^{\infty} \frac{(-1)^n d_n}{\frac{3}{2}(1-n)-s} \right), \quad (98)$$

$$g(2) = \lim_{t \to 0} t \left( \sum_{n=1}^{\infty} \frac{c_n}{n-1-t} + \frac{1}{\frac{3}{2}-2-t} + \sum_{n=1}^{\infty} \frac{(-1)^n d_n}{\frac{3}{2}(1-n)-2-t} \right). \quad (99)$$

Only the first term of the first series will survive the limit,

$$\Rightarrow g(2) = \lim_{t \to 0} t \left( \frac{c_1}{-t} \right) = -c_1 = \frac{2\pi}{3^{\frac{5}{6}} \Gamma\left(\frac{2}{3}\right)^2} = \frac{1}{\kappa}. \quad (100)$$

Similarly, for $s = 3$,

$$g(3) = \lim_{s \to 3} \frac{\sin(\pi s)}{\pi} \left( \sum_{n=1}^{\infty} \frac{c_n}{n+1-s} + \frac{1}{\frac{3}{2}-s} + \sum_{n=1}^{\infty} \frac{(-1)^n d_n}{\frac{3}{2}(1-n)-s} \right), \quad (101)$$

$$g(3) = \lim_{t \to 0} -t \left( \sum_{n=1}^{\infty} \frac{c_n}{n-2-t} + \frac{1}{-\frac{3}{2}-t} + \sum_{n=1}^{\infty} \frac{(-1)^n d_n}{\frac{3}{2}(-1-n)-t} \right). \quad (102)$$

The $n = 2$ term of the first series is the only one that will survive the limit,

$$\Rightarrow g(3) = -\lim_{t \to 0} t \left( \frac{c_2}{-t} \right) = c_2 = 1. \quad (103)$$

$s = 4$ produces the expected result,

$$g(4) = \lim_{s \to 4} \frac{\sin(\pi s)}{\pi} \left( \sum_{n=1}^{\infty} \frac{c_n}{n+1-s} + \frac{1}{\frac{3}{2}-s} + \sum_{n=1}^{\infty} \frac{(-1)^n d_n}{\frac{3}{2}(1-n)-s} \right), \quad (104)$$

$$g(4) = -c_3 = \frac{2\pi^2}{3^{\frac{5}{3}}\Gamma\left(\frac{2}{3}\right)^4} = \frac{1}{2\kappa^2}. \tag{105}$$

Note that there is a factor of 1/2 difference between this and Ref. 9, however this is the same result obtained in section III using the Weierstrass factorization. The above pattern repeats and can be expressed as,

$$g(n+1) = (-1)^n c_n \quad \text{for } n = 1, 2, 3, \ldots. \tag{106}$$

Next, values are computed for negative integers. For $s = -1$,

$$g(-1) = \lim_{s \to -1} \frac{\sin(\pi s)}{\pi}\left(\sum_{n=1}^{\infty}\frac{c_n}{n+1-s} + \frac{1}{\frac{3}{2}-s} + \sum_{n=1}^{\infty}\frac{(-1)^n d_n}{\frac{3}{2}(1-n)-s}\right), \tag{107}$$

$$g(-1) = \lim_{t \to 0} -t\left(\sum_{n=1}^{\infty}\frac{c_n}{n+2-t} + \frac{1}{\frac{5}{2}-t} + \sum_{n=1}^{\infty}\frac{(-1)^n d_n}{\frac{3}{2}(1-n)+1-t}\right), \tag{108}$$

$$g(-1) = 0 \quad \Rightarrow tr(H_\varepsilon) = 0. \tag{109}$$

For $s = -2$, the result is again zero,

$$g(-2) = \lim_{s \to -2} \frac{\sin(\pi s)}{\pi}\left(\sum_{n=1}^{\infty}\frac{c_n}{n+1-s} + \frac{1}{\frac{3}{2}-s} + \sum_{n=1}^{\infty}\frac{(-1)^n d_n}{\frac{3}{2}(1-n)-s}\right), \tag{110}$$

$$g(-2) = \lim_{t \to 0} t\left(\sum_{n=1}^{\infty}\frac{c_n}{n+3-t} + \frac{1}{\frac{7}{2}-t} + \sum_{n=1}^{\infty}\frac{(-1)^n d_n}{\frac{3}{2}(1-n)+2-t}\right), \tag{111}$$

$$g(-2) = 0. \tag{112}$$

For $s = -3$ a non-zero result is obtained, just as occurred for $f(-3)$,

$$g(-3) = \lim_{s \to -3} \frac{\sin(\pi s)}{\pi}\left(\sum_{n=1}^{\infty}\frac{c_n}{n+1-s} + \frac{1}{\frac{3}{2}-s} + \sum_{n=1}^{\infty}\frac{(-1)^n d_n}{\frac{3}{2}(1-n)-s}\right), \tag{113}$$

$$g(-3) = \lim_{t \to 0} -t\left(\sum_{n=1}^{\infty}\frac{c_n}{n+4-t} + \frac{1}{\frac{9}{2}-t} + \sum_{n=1}^{\infty}\frac{(-1)^n d_n}{\frac{3}{2}(1-n)+3-t}\right), \tag{114}$$

$$g(-3) = -d_3 = -\frac{21}{64}. \tag{115}$$

The above pattern continues, $g(-4) = g(-5) = 0$ and $g(-6) = -d_5$, etc.,

$$g(-3n) = (-1)^n d_{2n+1} \quad \text{for } n = 1,2,3 \ldots. \tag{116}$$

The above results are consistent with the results of Refs. 10 and 13 with one exception. Ref. 10 studied the Airy distribution via Laplace and Mellin transformations. As an off shoot of this study, they found that the Airy constants (which are proportional to the $b_n$, Eqs. (52) and (53) of this paper, were related to special values of the zeta function for the roots of the Airy function ($f(s)$ of this paper). They obtained values of the root zeta function at $s = 1, 0, -1, -2, \ldots$. This paper agrees with the results: $f(1), f(-1) = f(-2) = f(-4) = f(-5) = \cdots = 0$. It agrees with $f(-3n)$, but only up to a sign. Ref. 10 did not obtain results for $s > 1$. Ref. 13 studied the Schrödinger operator with potentials of the form $|z|^N$ using an exact WKB method. The results applicable to the Airy function ($N = 1$) gave results for the zeta functions for the odd and even states ($f(s)$ and $g(s)$ of this paper respectively) for $s = 0, 1, 2,$ and 3. This paper agrees with the results of Ref. 13. Ref. 13 did not obtain results for $s < 0$.

Computing special values for the slope is more challenging. The value of the slope at the origin is needed for the computation of the determinant. The derivative of $f(s)$, Eq. (61), is,

$$f'(s) = \cos(\pi s) \left( \sum_{n=0}^{\infty} \frac{a_n}{n+1-s} + \frac{1}{\frac{3}{2}-s} - \sum_{n=1}^{\infty} \frac{(-1)^n b_n}{\frac{3}{2}(1-n)-s} \right)$$
$$+ \frac{\sin(\pi s)}{\pi} \left( \sum_{n=0}^{\infty} \frac{a_n}{(n+1-s)^2} + \frac{1}{\left(\frac{3}{2}-s\right)^2} - \sum_{n=1}^{\infty} \frac{(-1)^n b_n}{\left(\frac{3}{2}(1-n)-s\right)^2} \right). \tag{117}$$

At the origin, the slope is given by the following limit,

$$f'(0) = \lim_{s \to 0} \left[ \cos(\pi s) \left( \sum_{n=0}^{\infty} \frac{a_n}{n+1-s} + \frac{1}{\frac{3}{2}-s} - \sum_{n=1}^{\infty} \frac{(-1)^n b_n}{\frac{3}{2}(1-n)-s} \right) \right.$$
$$\left. + \frac{\sin(\pi s)}{\pi} \left( \sum_{n=0}^{\infty} \frac{a_n}{(n+1-s)^2} + \frac{1}{\left(\frac{3}{2}-s\right)^2} - \sum_{n=1}^{\infty} \frac{(-1)^n b_n}{\left(\frac{3}{2}(1-n)-s\right)^2} \right) \right]. \tag{118}$$

Pulling terms from the limit where $s$ can unambiguously go to zero and deleting the terms that easily go to zero gives,

$$f'(0) = \sum_{n=0}^{\infty} \frac{a_n}{n+1} + \frac{2}{3} - \lim_{s \to 0} \left[ \sum_{n=1}^{\infty} \frac{(-1)^n b_n}{\frac{3}{2}(1-n) - s} + \sum_{n=1}^{\infty} \frac{s(-1)^n b_n}{\left(\frac{3}{2}(1-n) - s\right)^2} \right]. \tag{119}$$

Combining the two series within the limit into one yields,

$$f'(0) = \sum_{n=0}^{\infty} \frac{a_n}{n+1} + \frac{2}{3} - \lim_{s \to 0} \left[ \sum_{n=1}^{\infty} \frac{(-1)^n b_n \left(\frac{3}{2}(1-n)\right)}{\left(\frac{3}{2}(1-n) - s\right)^2} \right]. \tag{120}$$

The first term ($n = 1$) of the series within the limit is now seen to be zero. After letting $s \to 0$ and canceling common factors the slope at the origin is found to be,

$$f'(0) = \sum_{n=0}^{\infty} \frac{a_n}{n+1} + \frac{2}{3} - \frac{2}{3} \sum_{n=2}^{\infty} \frac{(-1)^n b_n}{(1-n)}. \tag{121}$$

This is a finite number. The first sum can be evaluated with an integral (which is easily computed with software such as MAPLE). Recall,

$$\sum_{n=0}^{\infty} a_n x^n = \frac{Ai'(x)}{Ai(x)}. \tag{122}$$

Now integrate Eq. (122) from zero to one,

$$\int_0^1 \sum_{n=0}^{\infty} a_n x^n dx = \int_0^1 \frac{Ai'(x)}{Ai(x)} dx, \tag{123}$$

$$\Rightarrow \sum_{n=0}^{\infty} \frac{a_n}{n+1} = \int_0^1 \frac{Ai'(x)}{Ai(x)} dx = -.96475832868595\ldots \tag{124}$$

The second sum is a bit harder to evaluate as the $b_n$ come from the series for $Ai'(x)/Ai(x)$ about $x = \infty$. It is, however, still possible to cast the result as an integral that is computable via software such as MAPLE. Recall,

$$\frac{\frac{d}{dx} Ai(x)}{Ai(x)} = \sqrt{x} \left( \sum_{n=1}^{\infty} (-1)^n b_n \left(x^{\frac{3}{2}}\right)^{-n} - 1 \right). \tag{125}$$

Let $x = y^{\frac{2}{3}}$, then Eq. (125) can be written as,

$$\sum_{n=2}^{\infty}(-1)^n b_n y^{-n} = \frac{\frac{d}{d\left(y^{\frac{2}{3}}\right)} Ai\left(y^{\frac{2}{3}}\right)}{y^{\frac{1}{3}} Ai\left(y^{\frac{2}{3}}\right)} + 1 + \frac{b_1}{y}. \qquad (126)$$

If this last equation is integrated between one and infinity the desired result is obtained,

$$-\sum_{n=2}^{\infty}\frac{(-1)^n b_n}{1-n} = \int_1^{\infty}\left(\frac{\frac{d}{d\left(y^{\frac{2}{3}}\right)} Ai\left(y^{\frac{2}{3}}\right)}{y^{\frac{1}{3}} Ai\left(y^{\frac{2}{3}}\right)} + 1 + \frac{1}{4y}\right) dy. \qquad (127)$$

The left-hand side is the series whose sum is needed, and the right-hand side is an integral that can be evaluated numerically with MAPLE,

$$\int_1^{\infty}\left(\frac{\frac{d}{d\left(y^{\frac{2}{3}}\right)} Ai\left(y^{\frac{2}{3}}\right)}{y^{\frac{1}{3}} Ai\left(y^{\frac{2}{3}}\right)} + 1 + \frac{1}{4y}\right) dy = .102207009191\ .... \qquad (128)$$

Substituting the numerical results, Eq. (124) and Eq. (128), into Eq. (121) gives the slope at the origin,

$$\Rightarrow f'(0) = -.2299536559\ .... \qquad (129)$$

The determinant of $H_o$ is then given by,

$$det(H_o) = \alpha^{-\frac{1}{4}} e^{(.2299536559...)} = \left(\frac{2\mu}{a^2\hbar^2}\right)^{\frac{1}{12}} e^{(.2299536559...)}. \qquad (130)$$

This result is now comparable with a result due to Voros[13]. Voros was able to compute the determinant without the physical factor, meaning $\alpha^{-\frac{1}{4}}$, using an exact WKB method,

$$\frac{det(H_o)}{\alpha^{-\frac{1}{4}}} = 2\sqrt{\pi} Ai(0) = \frac{2\sqrt{\pi}}{3^{2/3}\Gamma(2/3)} = 1.2585416826\ .... \qquad (131)$$

Here, using the contour integration method[11,12], it was found that,

$$\frac{det(H_o)}{\alpha^{-\frac{1}{4}}} = e^{(.2299536559...)} = 1.2585416826\ ..., \qquad (132)$$

which agrees to ten digits.

To compute the determinant for the even states the slope of $g(s)$ is needed at the origin. The derivative of $g(s)$, Eq. (90), is,

$$g'(s) = \cos(\pi s)\left(\sum_{n=1}^{\infty} \frac{c_n}{n+1-s} + \frac{1}{\frac{3}{2}-s} + \sum_{n=1}^{\infty} \frac{(-1)^n d_n}{\frac{3}{2}(1-n)-s}\right)$$
$$+ \frac{\sin(\pi s)}{\pi}\left(\sum_{n=1}^{\infty} \frac{c_n}{(n+1-s)^2} + \frac{1}{\left(\frac{3}{2}-s\right)^2} + \sum_{n=1}^{\infty} \frac{(-1)^n d_n}{\left(\frac{3}{2}(1-n)-s\right)^2}\right). \tag{133}$$

The slope at the origin is given by the following limit,

$$g'(0) = \lim_{s \to 0}\left[\cos(\pi s)\left(\sum_{n=1}^{\infty} \frac{c_n}{n+1-s} + \frac{1}{\frac{3}{2}-s} + \sum_{n=1}^{\infty} \frac{(-1)^n d_n}{\frac{3}{2}(1-n)-s}\right)\right.$$
$$\left. + \frac{\sin(\pi s)}{\pi}\left(\sum_{n=1}^{\infty} \frac{c_n}{(n+1-s)^2} + \frac{1}{\left(\frac{3}{2}-s\right)^2} + \sum_{n=1}^{\infty} \frac{(-1)^n d_n}{\left(\frac{3}{2}(1-n)-s\right)^2}\right)\right], \tag{134}$$

Pulling terms from the limit where $s$ can unambiguously go to zero and deleting the terms that easily go to zero gives,

$$g'(0) = \sum_{n=1}^{\infty} \frac{c_n}{n+1} + \frac{2}{3} + \lim_{s \to 0}\left[\sum_{n=1}^{\infty} \frac{(-1)^n d_n}{\frac{3}{2}(1-n)-s} + \sum_{n=1}^{\infty} \frac{s(-1)^n d_n}{\left(\frac{3}{2}(1-n)-s\right)^2}\right]. \tag{135}$$

Now combine the two series within the limit,

$$g'(0) = \sum_{n=1}^{\infty} \frac{c_n}{n+1} + \frac{2}{3} + \lim_{s \to 0}\left[\sum_{n=1}^{\infty} \frac{\frac{3}{2}(1-n)(-1)^n d_n}{\left(\frac{3}{2}(1-n)-s\right)^2}\right]. \tag{136}$$

The first term ($n = 1$) of the series within the limit is now seen to be zero. This leaves the following result for the slope at the origin,

$$g'(0) = \sum_{n=1}^{\infty} \frac{c_n}{n+1} + \frac{2}{3} + \lim_{s \to 0}\left[\sum_{n=2}^{\infty} \frac{\frac{3}{2}(1-n)(-1)^n d_n}{\left(\frac{3}{2}(1-n)-s\right)^2}\right]. \tag{137}$$

After letting $s \to 0$ and canceling common factors the slope at the origin is found to be,

$$g'(0) = \sum_{n=1}^{\infty} \frac{c_n}{n+1} + \frac{2}{3} + \frac{2}{3}\sum_{n=2}^{\infty} \frac{(-1)^n d_n}{(1-n)}. \tag{138}$$

The two sums can be evaluated in the same way as the case for the odd states was handled,

$$\sum_{n=1}^{\infty} \frac{c_n}{n+1} = \int_0^1 \frac{Ai''(x)}{Ai'(x)} dx = -0.4862994595385353 \ldots \tag{139}$$

The second sum, in Eq. (138), requires a few more steps. Recall,

$$\frac{Ai''(x)}{Ai'(x)} = -\sqrt{x} - \sqrt{x} \sum_{n=1}^{\infty} (-1)^n d_n \left(x^{\frac{3}{2}}\right)^{-n}. \tag{140}$$

Now let $x = y^{\frac{2}{3}}$ and rearrange to obtain,

$$-\sum_{n=2}^{\infty} (-1)^n d_n y^{-n} = \frac{\frac{d^2}{d(y^{\frac{2}{3}})^2} Ai(y^{\frac{2}{3}})}{\sqrt[3]{y} \frac{d}{d(y^{\frac{2}{3}})} Ai(y^{\frac{2}{3}})} + 1 - \frac{1}{4y}. \tag{141}$$

Integrating Eq. (141) from one to infinity yields the desired result,

$$\sum_{n=2}^{\infty} \frac{(-1)^n d_n}{1-n} = \int_1^{\infty} \left( \frac{\frac{d^2}{d(y^{\frac{2}{3}})^2} Ai(y^{\frac{2}{3}})}{\sqrt[3]{y} \frac{d}{d(y^{\frac{2}{3}})} Ai(y^{\frac{2}{3}})} + 1 - \frac{1}{4y} \right) dy, \tag{142}$$

The left-hand side is the series whose sum is needed, and the right-hand side is an integral that can be evaluated numerically with MAPLE,

$$\Rightarrow \sum_{n=2}^{\infty} \frac{(-1)^n d_n}{1-n} = -.141381881194 \ldots \tag{143}$$

Substituting the numerical results, Eq. (139) and Eq. (143), into Eq. (138) gives the slope at the origin,

$$\Rightarrow g'(0) = .08611261973\ldots \tag{144}$$

$$det(H_\varepsilon) = \alpha^{\frac{1}{4}} e^{-.08611261973\ldots} = \left(\frac{a^2\hbar^2}{2\mu}\right)^{\frac{1}{12}} e^{-.08611261973\ldots}. \tag{145}$$

Note that,
$$e^{-.08611261973\ldots} = .9174908978\ldots \tag{146}$$

Voros[13] found

$$\frac{det(H_\varepsilon)}{\alpha^{\frac{1}{4}}} = -2\sqrt{\pi} Ai'(0) = .91749089788326\ldots \tag{147}$$

which agrees to 9 digits. Note also that there was no need for normalization to obtain the determinants in the present calculation (by contrast see the discussion below Eq. (27) of Ref. 13).

## V. CONCLUSION

In this paper, properties of the zeta functions for the triangular potential were investigated. Values of the zeta functions were computed for integers greater than one using both the Weierstrass factorization theorem and analytic continuation via contour integration and found to be consistent. Analytic continuation was used to compute values of the zeta functions at zero and the negative integers as well as the value of the slopes at the origin. This allowed for the computation of the trace and determinant of the associated Hamiltonians. To summarize, the following results were found,

$$\zeta_{H_o}(s) = \left(\frac{a^2\hbar^2}{2\mu}\right)^{-\frac{s}{3}} \frac{sin(\pi s)}{\pi} \left(\sum_{n=0}^{\infty} \frac{a_n}{n+1-s} + \frac{1}{\frac{3}{2}-s} - \sum_{n=1}^{\infty} \frac{(-1)^n b_n}{\frac{3}{2}(1-n)-s}\right), \tag{148}$$

$$\zeta_{H_o}(n+1) = (-1)^n \left(\frac{a^2\hbar^2}{2\mu}\right)^{-\frac{n+1}{3}} a_n \quad \text{for } n = 0,1,2,\ldots, \tag{149}$$

$$\zeta_{H_o}(0) = -\frac{1}{4}, \tag{150}$$

$$\zeta_{H_o}(-3n) = (-1)^n \left(\frac{a^2\hbar^2}{2\mu}\right)^n (-1)^n b_{2n+1} \quad \text{for } n = 1,2,3,\ldots, \tag{151}$$

for $s$ being a negative integer that is not divisible by 3,

$$\zeta_{H_o}(s) = 0, \tag{152}$$

$$\zeta_{H_\varepsilon}(s) = \left(\frac{a^2\hbar^2}{2\mu}\right)^{-\frac{s}{3}} \frac{sin(\pi s)}{\pi} \left(\sum_{n=1}^{\infty} \frac{c_n}{n+1-s} + \frac{1}{\frac{3}{2}-s} + \sum_{n=1}^{\infty} \frac{(-1)^n d_n}{\frac{3}{2}(1-n)-s}\right), \quad (153)$$

$$\zeta_{H_\varepsilon}(n+1) = (-1)^n \left(\frac{a^2\hbar^2}{2\mu}\right)^{-\frac{s}{3}} c_n \quad \text{for } n = 1,2,3,\ldots, \quad (154)$$

$$\zeta_{H_\varepsilon}(0) = \frac{1}{4}, \quad (155)$$

$$\zeta_{H_\varepsilon}(-3n) = (-1)^n \left(\frac{a^2\hbar^2}{2\mu}\right)^{-\frac{s}{3}} d_{2n+1} \quad \text{for } n = 1,2,3\ldots, \quad (156)$$

for $s$ being a negative integer that is not divisible by 3,

$$\zeta_{H_\varepsilon}(s) = 0. \quad (157)$$

The zeta function for the full Hamiltonian, where the odd and even states are combined, can be obtained rather trivially by adding the two zeta functions together,

$$\zeta_H(s) = \left(\frac{a^2\hbar^2}{2\mu}\right)^{-\frac{s}{3}} \frac{sin(\pi s)}{\pi} \left(\sum_{n=0}^{\infty} \frac{a_n + c_n}{n+1-s} + \frac{2}{\frac{3}{2}-s} - \sum_{n=1}^{\infty} \frac{(-1)^n(b_n + d_n)}{\frac{3}{2}(1-n)-s}\right). \quad (158)$$

It is interesting that the determinant for the full zeta function is a number with no physical dimensions,

$$det(H) = \alpha^{\frac{1}{4}-\frac{1}{4}} e^{.2299536559\ldots - .08611261973\ldots} = 1.154700538\ldots, \quad (159)$$

the exponent due to the odd states exactly cancels with the exponent due to the even states.

Computationally it is interesting to note that to do the zeta function calculations the actual values of the roots of the Airy function, or its slope, were not needed. All that is needed are the coefficients of the series expansions of the logarithmic derivative of the Airy function and its derivative about the origin and infinity. The coefficients for the series representation of the logarithmic derivative of the Airy function about the origin giving the values of the zeta functions at the positive integers, the odd indexed coefficients of the series expansion about infinity giving the values of the zeta functions at the negative integers, and the even indexed coefficients giving the residues.

**DATA AVAILABILITY**

Data sharing is not applicable to this article as no new data were created or analyzed in this study.

**REFERENCES**

1. [H] J. Schwinger, *Quantum Mechanics, Symbolism of Atomic Measurements*, edited by B-G. Englert, (Springer 2001).

2. [I] L.D. Landua and E.M. Lifshitz, *Quantum Mechanics (Non-relativistic Theory)*, 3rd edition (Pergamon, 1977).

3. [G] R. E. Crandall, *On the quantum zeta function*, J. Phys. A: Math. Gen. 29, 6795-6816 (1996). DOI: 10.1088/0305-4470/29/21/014.

4. [J] E. Elizalde, S.D. Odintsov, A. Romeo, A.A. Bytsenk, and S. Zerbini, *Zeta Regularization Techniques with Applications*, World Scientific (1994).

5. [K] E. Elizalde, *Ten Physical Applications of Spectral Zeta Functions*, 2nd ed., Lecture Notes in Physics 855, Springer-Verlag (2012).

6. [L] K. Kirsten, *Spectral functions in mathematics and physics*, Chapman&Hall/CRC, Boca Raton, FL, (2002).

7. [D] K. Kirsten, *Functional determinants in higher dimensions using contour integrals*, in *A Window into Zeta and Modular Physics* (Mathematical Sciences Research Institute Publications) Edited by K. Kirsten and F. L. Williams, 1st ed., Cambridge University Press (2010).

8. [M] L. A. Rubel with J. E. Colliander, *Entire and Meromorphic Functions*, Springer (1996).

9. [A] O. Vallée and M. Soares, *Airy Functions and Applications to Physics* 2nd ed., World Scientific (2010).

10. [E] P. Flajolet and G. Louchard, *ANALYTIC VARIATIONS ON THE AIRY DISTRIBUTION*, Algorithmica, 31, 361-377 (2001). https://doi.org/10.1007/s00453-001-0056-0.

11. [B] K. Kirsten and P. Loya, *Computation of determinants using contour integrals*, Am. J. Phys. 76: 60-64, 2008, DOI: 10.1119/1.2794348. Also available at arXiv:0707.3755 [hep-th].

12. [C] E. Elizalde, K. Kirsten, N. Robles, and F. Williams, *Zeta functions on tori using contour integration*, Int. J. Geom. Methods Mod. Phys. 12, 1550019 (2015). DOI: 10.1142/S021988781550019X. Also available at arXiv:1306.4019 [math-ph].

13. A. Voros, *Airy function (exact WKB results for potentials of odd degree)*, J. Phys. A: Math. Gen. **32**, No. 7, 1301 (1999). DOI: 10.1088/0305-4470/32/7/020. Also available at, arXiv:math-ph/9811001v2